\documentclass[twocolumn,twoside,slac_two]{revtex4}
\usepackage{graphicx}
\usepackage{fancyhdr}
\begin{document}

\title{Study for relation between direction of relativistic jet
and optical polarization angle with multi-wavelength observation}

%

\author{R.~Itoh, Y.~Fukazawa, K.~Kawaguchi, Y.~Kanda}
\affiliation{Department of Physical Science, Hiroshima University, Higashi-Hiroshima, Hiroshima 739-8526, Japan}
\author{Y.~T.~Tanaka, M.~Uemura, K.~S.~Kawabata, H.~Akitaya}
\affiliation{Hiroshima Astrophysical Science Center, Hiroshima University, Higashi-Hiroshima, Hiroshima 739-8526, Japan}
\author{for the Kanata and OISTER team}

\begin{abstract}
Blazars are thought to possess a relativistic jet that is
pointing toward the direction of the Earth and the elect of
relativistic beaming enhances its apparent brightness.
They radiate in all wavebands from the radio to the gamma-ray
bands via the synchrotron and the inverse Compton scattering process.
Numerous observations are performed but the mechanism of
variability, creation and composition of jets are still controversial.

We performed multi-wavelength monitoring with optical polarization
for 3C~66A, Mrk~421, CTA~102 and PMN~J0948+0022 to investigate
the mechanisms of variability and research the emission region in
the relativistic jets.
Consequently, an emergence of new emission component in flaring
state is suggested in each object.
The most significant aspect of these results is its wide range of sizes
of emission regions from $10^{14}-10^{16}$ cm, which implies the model
with a number of independent emission regions with variety sizes and
randomly orientation.
The "shock-in-jet" scenario can explain high PD and direction of PA
in each objects.
It might reflect the common mechanism of flares in the relativistic jets.
\end{abstract}

\maketitle

\thispagestyle{fancy}


\section{INTRODUCTION}

Blazars are highly variable active galactic nuclei (AGN) which emit radiation
at all wavelengths from radio to gamma-rays. 
They have strong relativistic jets aligned with the observer's line of sight 
and are apparently bright due to relativistic beaming.
Outstanding characteristics of blazars are their rapid and
high-amplitude intensity variations or flares. 
Blazar consists of several sub-classes.
BL Lac objects are detected to have weak emission line of equivalent width 
$< 5\ $\AA\ in the observer's optical band definition.
In contrast, flat spectrum radio quasars (FSRQs) shows relatively strong emission lines. 
Blazars also can be classified into three types, 
based on their peak frequency of synchrotron radiation $\nu_{\rm peak}^{S}$ \citep{2011ApJ...743..171A};
low-synchrotron-peaked blazars (LSP; for sources with $\nu_{\rm peak}^{S} < 10^{14}$ Hz), 
intermediate-synchrotron-peaked blazars (ISP; for $10^{14} {\ \rm Hz} < \nu_{\rm peak}^{S} < 10^{15}$ Hz) 
and high-synchrotron-peaked blazars (HSP; for $10^{15} {\ \rm Hz} < \nu_{\rm peak}^{S}$).
Due to relativistic effect, radiation from jets dominates the overall spectral energy distribution 
and hence, their spectra in the optical band are featureless compared with other AGNs. 
From this reason, blazar is one of most suitable objects to study the jets.

Polarized radiation from blazars is one of the evidence of synchrotron radiation 
in low energies and it also varies drastically.
The polarization of blazars is of interest for understanding the origin, confinement, and propagation of jets 
\citep{1985agn..book..215B,1998AJ....116.2119V}.
Mead et al. (1990)\citep{1990A&AS...83..183M} performed a large-sample study of blazars in the optical band 
and showed that high polarization degree (PD) and variability of polarization are common phenomena in blazars.  
Ikejiri et al. (2011)\citep{2011PASJ...63..639I} reported statistical photopolarimetric observations of blazars with a daily timescale, 
and suggested that lower luminosity and higher 
peak frequency of synchrotron radiation objects (such as HSP blazar) had 
smaller amplitudes in their variations both in the flux, color, and PD.
The author also reported the about 30\% of blazars showed correlation between the optical flux and PD.
Numerous observations are performed but the mechanism of variability, 
creation and composition of jets are still controversial.

In efforts to find a common mechanism of jets, 
we performed observations on various types of AGNs.
Simultaneous multi-wavelength and optical polarimetric observations 
are powerful tools to probe the emission region in jets, 
thus we performed wide-band multi-wavelength (from radio to TeV gamma-ray) 
observations of relativistic jets in several types of AGNs with various 
timescales (from minute to year) to study of structures and emission regions of relativistic jets.

\section{Observations}

We constructed the framework of multi-wavelength and optical polarimetric observations of 
relativistic jets in AGNs with the {\it Fermi} Gamma-ray Space Telescope,
Monitor of All-sky X-ray Image \citep[MAXI;][]{2009PASJ...61..999M},
the {\it Swift} Gamma-Ray-Burst Explorer \citep{2004ApJ...611.1005G},
the Kanata optical and near infrared telescope, 
Optical and Infrared Synergetic Telescopes for Education and Research ({\it OISTER}),
and Mizusawa VLBI Observatory.

We performed four objects observation with optical flux and polarization  
to see the relations between polarization angle (PA) and the direction
of radio jets in the flaring state.
In efforts to find a common mechanism of jets, 
we selected the different types of AGNs to see the difference 
between them.
This study focused on four AGNs; ISP blazar 3C~66A,
HSP blazar Mrk~421, FSRQ CTA~102 and radio-loud narrow-line Seyfert 1 galaxy (RL-NLSy1) PMN~J0948+0022.
Note that RL-NLSy1 is not the class of blazar but it thought to possess relativistic jet \citep{2009ApJ...707L.142A}.
Some radio galaxies also known as GeV gamma-ray emitter and those class of AGNs might play an important role
to probe the emission region in the relativistic jets (e.g., \citep{2015ApJ...799L..18T}).
Individual results are reported in Itoh et~al. (2013a)\citep{2013PASJ...65...18I}, 
Itoh et~al. (2015, submitted to the PASJ)\citep{2015mkn421},
Itoh et~al. (2013b)\citep{2013ApJ...768L..24I} and Itoh et~al. (2013c)\citep{2013ApJ...775L..26I}
respectively.
We selected the flare with good correlation between polarized flux 
and total flux in optical band.

\section{Summary of case studies}

\begin{figure*}[!htb]
 \centering
 \includegraphics[angle=0,width=8cm]{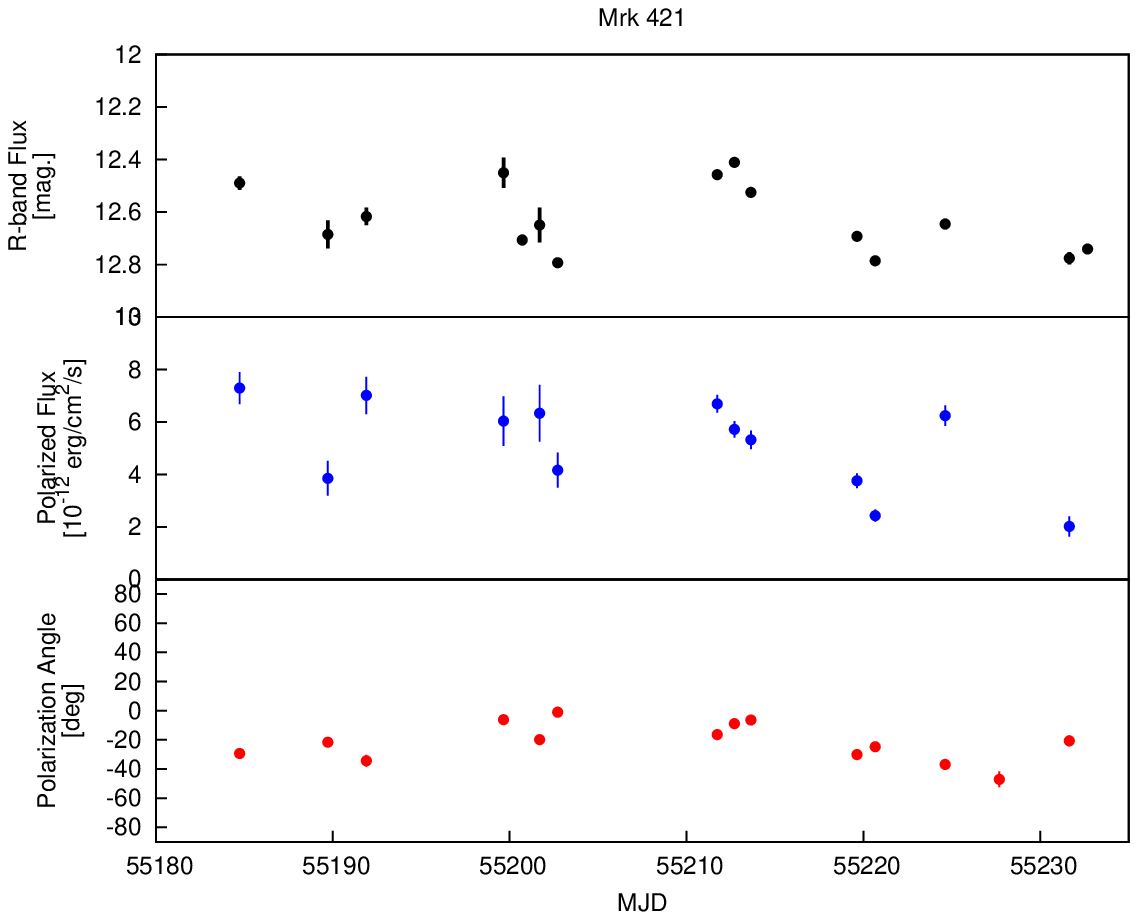}
 \includegraphics[angle=0,width=8cm]{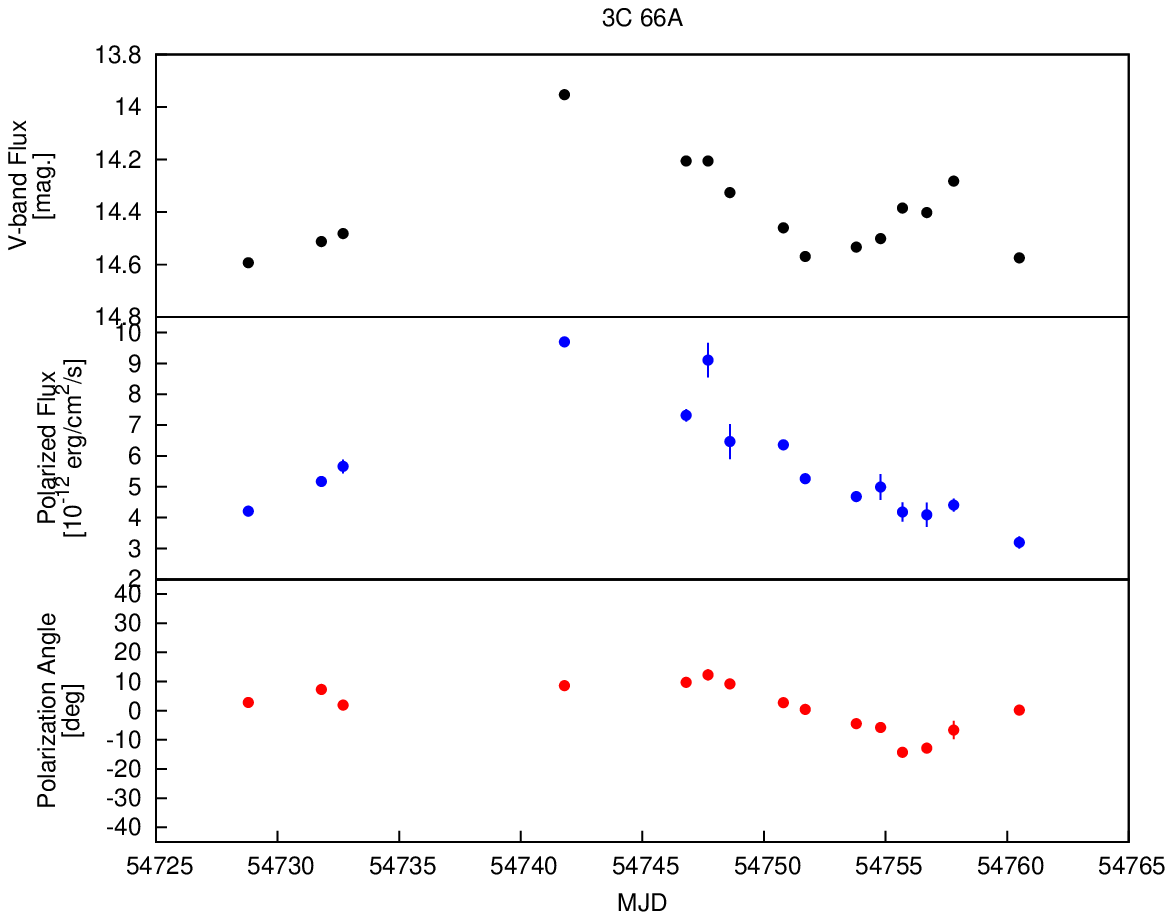}
 \includegraphics[angle=0,width=8cm]{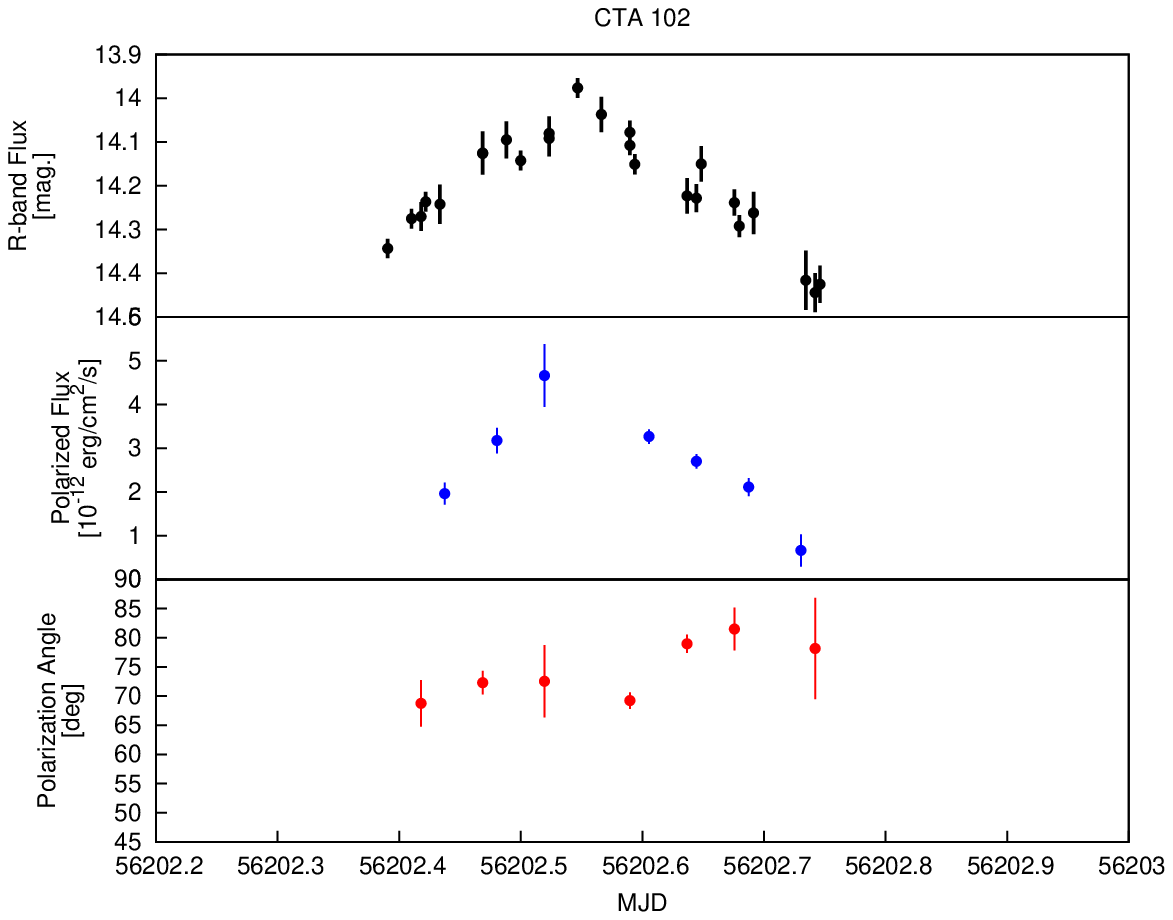}
 \includegraphics[angle=0,width=8cm]{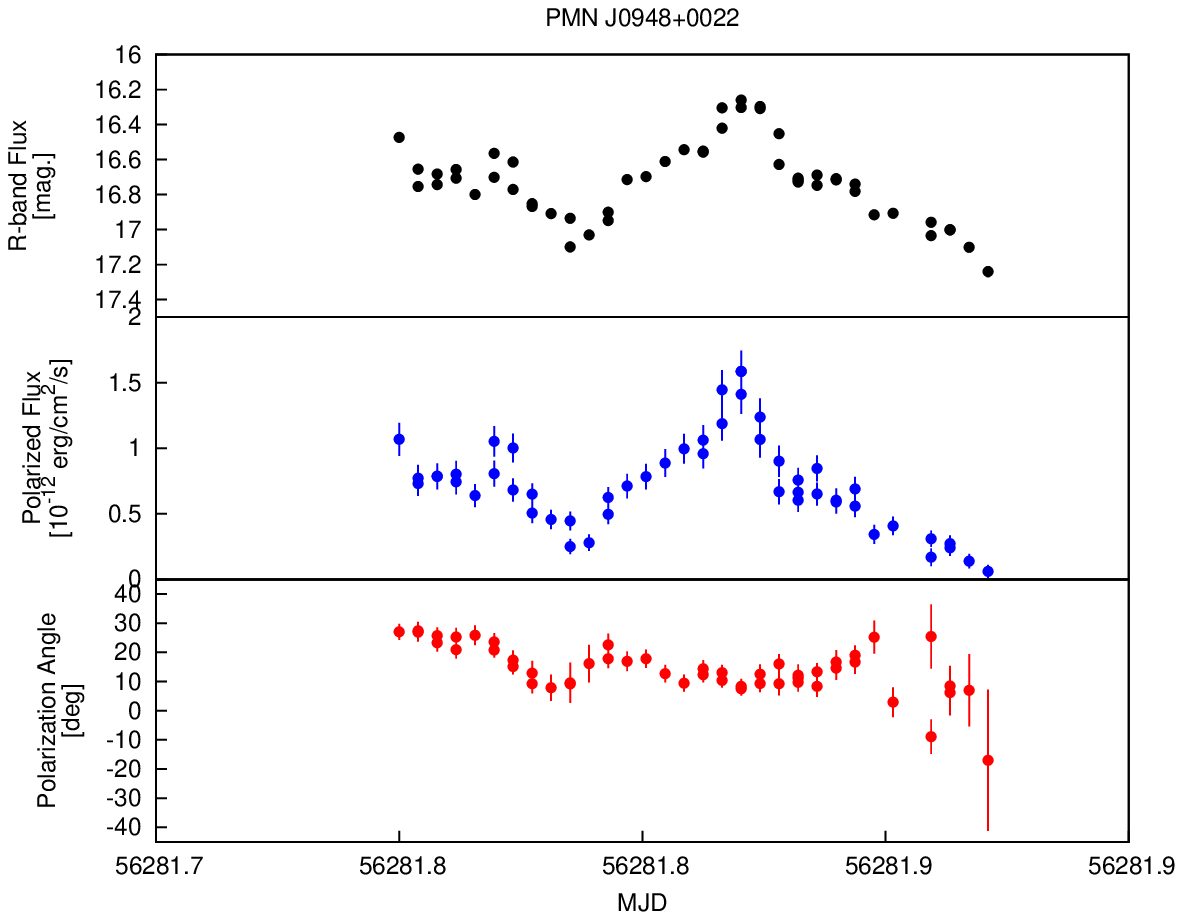}
 \caption{Light curves for each source. From top to bottom, 
   the histories of the total flux in the $R_{C}$ band, the polarized flux,
   and the polarization angle (PA) are shown.
   Note that the time scale of CTA~102 and PMN~J0948+0022 (bottom two sources) 
   are different from that in 3C~66A and Mrk~421.
   Details of each light curve are reported in \cite{2013PASJ...65...18I}, 
   \cite{2013ApJ...768L..24I}, \cite{2013ApJ...775L..26I} and \cite{2015mkn421}}
 \label{fig:LC}
\end{figure*}

In this section, we summarize our studies of individual blazar, 
Temporal variability in optical flux, polarization flux and PA
 are shown in figure \ref{fig:LC}.
Table \ref{tab:PA} shows a summary of differential angle ($\Delta$DA) between position angle of 
radio jet measured by VLBI or VLBA (\cite{2007A&A...468..963C}, \cite{2005ApJ...622..168P}, 
\cite{2013A&A...551A..32F} and \cite{2006PASJ...58..829D})

\paragraph{3C~66A}
We studied the long-term variations of 3C~66A 
over 2 years in the GeV band with {\it Fermi} 
and in the optical (flux and polarization) and near infrared band 
with the Kanata telescope.  
In 2008, we find a correlation between
the gamma-ray flux and the optical properties.  
This is in contrast to the later behaviours during 
2009--2010, a weak correlation along with a gradual
increase of the optical flux. 
We conclude that the different behaviors observed between the first
year and the later years might be explained by postulating two
different emission components. 
$\Delta$DA shown in Table \ref{tab:PA} indicates that the position angle of 
radio jet is close to the average PA.
It should be noted that a correlation between PD and total flux 
is significant in 2009.

\paragraph{Mrk~421}
We observed the long-term variability of Mrk 421 from optical to 
X-ray band using the {\it Swift}, MAXI, and Kanata telescope from 2010 to 2011.
In 2010, the variability in the X-ray band is clearly large, while the optical
and UV flux shows gradual decreasing.
Polarization properties also show the unique variability in 2010.
The variation on the Stokes parameter QU plane suggested the presence of the proper polarization.
On the other hand, the variability in the X-ray band is small in 2011,
although the variability in the optical and UV band is relativistically
large compared with that in 2010.
We speculated that Mrk 421 has different variability mechanisms between 
2010 and 2011 and emergence of a new emission component which have 
systematic difference of polarization at different periods.
$\Delta$DA  indicates that 
PA is aligned to the parsec scale jet in 2010.
We also found a good correlation between optical flux and polarized flux in 2010.
These behaviours are similar to that in 3C~66A.

\paragraph{CTA~102}
We densely monitored CTA~102 in the optical
and near-infrared bands for the subsequent ten nights using {\it OISTER},
following {\it Fermi}-LAT detection of the enhanced gamma-ray activity. 
On MJD 56197 (2012 September 27, 4-5 days after the peak of bright
gamma-ray flare), a polarized flux showed a transient increase, while
a total flux and PA remained almost constant during the
``orphan polarized-flux flare''.
We also detected an intra-night and prominent flare on MJD 56202. 
Emergence of a new emission component with high PD up to 40\% 
would be responsible for the observed two flares, and
such a high PD indicates a presence of highly ordered magnetic field
at the emission site. 
The observed directions of PA is perpendicular to the jet.
The total and polarized fluxes showed quite similar temporal
variations, but PA again remained constant during the flare. 

\paragraph{PMN~J0948+0022}
We performed optical photopolarimetric monitoring of the
RL-NLSy1 galaxy PMN~J0948+0022 on 2012 December to 2013 February
triggered by the flux enhancement in near infrared and $\gamma$-ray bands.
Thanks to one-shot polarimetry of the HOWPol installed to the Kanata
telescope, we have detected very rapid variability in the
polarized-flux light curve on MJD 56281 (2012 December 20).
The rise and decay times were about 140 sec and 180 sec, respectively.
The PD reached $36 \pm 3$\% at the peak of the
short-duration pulse, while PA remained almost
constant.
The high PD provides a clear evidence of synchrotron emission within a
highly ordered magnetic field at the emission site.  
These results provide new observational evidence that highly ordered
magnetic field is present inside a very compact emission region of the
order of $\sim10^{14}$ cm and imposes severe constraint on theoretical
studies unless central black hole mass is much smaller than currently
considered.
we found that PA in MJD 56202 is aligned to the parsec scale jet.
Temporal profiles of the total flux and PD showed highly
variable but well correlated behavior and discrete correlation
function analysis revealed that no significant time lag of more than
10 min was present.

\begin{table}[t]
\begin{center}
\caption{Summary of differential angle for each object.}
\begin{tabular}{llc}
\hline \textbf{Object name} & \textbf{AGN type} &\textbf{{\it $\Delta$DA}}${}^1$ [deg]\\ \hline 
3C~66A         & BL Lac (ISP) & $0 \pm 5 $\\
Mrk~421        & BL Lac (HSP) & $10\pm 10$ \\
CTA~102        & FSRQ       & $80\pm 10$ \\
PMN J0948+0022 & RL-NLSy1   & $5 \pm 5$ \\
\hline
\end{tabular}
\label{tab:PA}
\begin{flushleft} 
 ${}^1$: Differential angle between the position angle of radio jet and optical polarization angle.
\end{flushleft}
\end{center}
\end{table}

\section{Discussion} 
A common characteristic among BL lac objects and RL-NLSy1 is that
PA aligned with a direction parallel to the jet 
(see Table\ref{tab:PA}).
This phenomenon is well explained with the framework of
``shock-in-jet'' scenario, 
in which high PD and direction of PA are well explained with 
compressed emission region by the internal shocks.
This phenomena is explained with below mechanism; 
a compressed shock that is perpendicular to the jet flow
results that the electric polarization vector to be perpendicular 
to the emission blob and aligned with the jet axis.
Impey et~al. (2011)\citep{1991ApJ...375...46I} reported that about 60\% quasars shows alignment of the position
angle of jet and polarization angle.
Especially, author found that in 10 out of 11 BL Lac objects shows good alignment.
It should be noted that these measurements of PA were collected 
without considering the flux state. 
Similar tendencies in hourly-scale variability were reported in 
other BL Lac objects \citep[e.g., AO 0235+164,][]{2008ApJ...672...40H}.
On the other hand, CTA 102 which is classified as FSRQ shows 
a different tendency.
The difference of relation between PA and direction of the jet 
might be reflecting a difference of jets between BL lac objects and FSRQs.
In general, FSRQs thought to have weaker shocks and/or a stronger underlying 
magnetic fields such as large-scale helical magnetic fields.
Given this complicated situation, the measured PAs significantly 
different from the jet direction can still be accounted for by the ``shock-in-jet" scenario.
Therefore, it is suggested that the ``shock-in-jet'' is a common phenomena in 
relativistic jets, which independent on the synchrotron peaks, types of AGNs and timescale.
Similar relations between PA and direction of the radio jet 
are reported in measurements of radio polarization \citep{1998ApJ...504..702L}.
It might reflect the common mechanism of flares in the relativistic jets but 
we need more sample to confirm this trend.

\section{Acknowledgments}
This work is supported by Optical \& Near-infrared Astronomy Inter-University
Cooperation Program and Grants-in-Aid for Scientific Research 
(23340048, 24000004, 24244014, and 24840031) by the
Ministry of Education, Culture, Sports, Science and Technology of Japan.

\bigskip 

\end{document}